\documentstyle[twoside]{article}

\catcode`\@=11
\long\def\@makefntext#1{
\protect\noindent \hbox to 3.2pt {\hskip-.9pt  
$^{{\eightrm\@thefnmark}}$\hfil}#1\hfill}		

\def\thefootnote{\fnsymbol{footnote}}
\def\@makefnmark{\hbox to 0pt{$^{\@thefnmark}$\hss}}	
	
\def\ps@myheadings{\let\@mkboth\@gobbletwo
\def\@oddhead{\hbox{}
\rightmark\hfil\eightrm\thepage}   
\def\@oddfoot{}\def\@evenhead{\eightrm\thepage\hfil
\leftmark\hbox{}}\def\@evenfoot{}
\def\sectionmark##1{}\def\subsectionmark##1{}}

\oddsidemargin=\evensidemargin
\renewcommand{\thefootnote}{\fnsymbol{footnote}}

\newcounter{sectionc}\newcounter{subsectionc}\newcounter{subsubsectionc}
\renewcommand{\section}[1] {\vspace{12pt}\addtocounter{sectionc}{1} 
\setcounter{subsectionc}{0}\setcounter{subsubsectionc}{0}\noindent 
	{\tenbf\thesectionc. #1}\par\vspace{5pt}}
\renewcommand{\subsection}[1] {\vspace{12pt}\addtocounter{subsectionc}{1} 
	\setcounter{subsubsectionc}{0}\noindent 
	{\bf\thesectionc.\thesubsectionc. {\kern1pt \bfit #1}}\par\vspace{5pt}}
\renewcommand{\subsubsection}[1] {\vspace{12pt}\addtocounter{subsubsectionc}{1}
	\noindent{\tenrm\thesectionc.\thesubsectionc.\thesubsubsectionc.
	{\kern1pt \tenit #1}}\par\vspace{5pt}}
\newcommand{\nonumsection}[1] {\vspace{12pt}\noindent{\tenbf #1}
	\par\vspace{5pt}}

\newcounter{appendixc}
\newcounter{subappendixc}[appendixc]
\newcounter{subsubappendixc}[subappendixc]
\renewcommand{\thesubappendixc}{\Alph{appendixc}.\arabic{subappendixc}}
\renewcommand{\thesubsubappendixc}
	{\Alph{appendixc}.\arabic{subappendixc}.\arabic{subsubappendixc}}

\renewcommand{\appendix}[1] {\vspace{12pt}
        \refstepcounter{appendixc}
        \setcounter{figure}{0}
        \setcounter{table}{0}
        \setcounter{lemma}{0}
        \setcounter{theorem}{0}
        \setcounter{corollary}{0}
        \setcounter{definition}{0}
        \setcounter{equation}{0}
        \renewcommand{\thefigure}{\Alph{appendixc}.\arabic{figure}}
        \renewcommand{\thetable}{\Alph{appendixc}.\arabic{table}}
        \renewcommand{\theappendixc}{\Alph{appendixc}}
        \renewcommand{\thelemma}{\Alph{appendixc}.\arabic{lemma}}
        \renewcommand{\thetheorem}{\Alph{appendixc}.\arabic{theorem}}
        \renewcommand{\thedefinition}{\Alph{appendixc}.\arabic{definition}}
        \renewcommand{\thecorollary}{\Alph{appendixc}.\arabic{corollary}}
        \renewcommand{\theequation}{\Alph{appendixc}.\arabic{equation}}
        \noindent{\tenbf Appendix \theappendixc #1}\par\vspace{5pt}}
\newcommand{\subappendix}[1] {\vspace{12pt}
        \refstepcounter{subappendixc}
        \noindent{\bf Appendix \thesubappendixc. {\kern1pt \bfit #1}}
	\par\vspace{5pt}}
\newcommand{\subsubappendix}[1] {\vspace{12pt}
        \refstepcounter{subsubappendixc}
        \noindent{\rm Appendix \thesubsubappendixc. {\kern1pt \tenit #1}}
	\par\vspace{5pt}}

\newcommand{\textlineskip}{\baselineskip=13pt}
\newcommand{\smalllineskip}{\baselineskip=10pt}

\def\eightcirc{
\begin{picture}(0,0)
\put(4.4,1.8){\circle{6.5}}
\end{picture}}
\def\eightcopyright{\eightcirc\kern2.7pt\hbox{\eightrm c}} 

\newcommand{\copyrightheading}[1]
	{\vspace*{-2.5cm}\smalllineskip{\flushleft
	{\footnotesize International Journal of Modern Physics A, #1}\\
	{\footnotesize $\eightcopyright$\, World Scientific Publishing
	 Company}\\
	 }}

\def\abstracts#1#2#3{{
	\centering{\begin{minipage}{4.5in}\baselineskip=10pt\footnotesize
	\parindent=0pt #1\par 
	\parindent=15pt #2\par
	\parindent=15pt #3
	\end{minipage}}\par}} 

\newcommand{\bibit}{\nineit}

\renewenvironment{thebibliography}[1]
	{\frenchspacing
	 \ninerm\baselineskip=11pt
	 \begin{list}{\arabic{enumi}.}
	{\usecounter{enumi}\setlength{\parsep}{0pt}
	 \setlength{\leftmargin 12.7pt}{\rightmargin 0pt} 
	 \setlength{\itemsep}{0pt} \settowidth
	{\labelwidth}{#1.}\sloppy}}{\end{list}}

\newcounter{itemlistc}
\newcounter{romanlistc}
\newcounter{alphlistc}
\newcounter{arabiclistc}

\newcommand{\fcaption}[1]{
        \refstepcounter{figure}
        \setbox\@tempboxa = \hbox{\footnotesize Fig.~\thefigure. #1}
        \ifdim \wd\@tempboxa > 5in
           {\begin{center}
        \parbox{5in}{\footnotesize\smalllineskip Fig.~\thefigure. #1}
            \end{center}}
        \else
             {\begin{center}
             {\footnotesize Fig.~\thefigure. #1}
              \end{center}}
        \fi}

\newcommand{\tcaption}[1]{
        \refstepcounter{table}
        \setbox\@tempboxa = \hbox{\footnotesize Table~\thetable. #1}
        \ifdim \wd\@tempboxa > 5in
           {\begin{center}
        \parbox{5in}{\footnotesize\smalllineskip Table~\thetable. #1}
            \end{center}}
        \else
             {\begin{center}
             {\footnotesize Table~\thetable. #1}
              \end{center}}
        \fi}

\def\@citex[#1]#2{\if@filesw\immediate\write\@auxout
	{\string\citation{#2}}\fi
\def\@citea{}\@cite{\@for\@citeb:=#2\do
	{\@citea\def\@citea{,}\@ifundefined
	{b@\@citeb}{{\bf ?}\@warning
	{Citation `\@citeb' on page \thepage \space undefined}}
	{\csname b@\@citeb\endcsname}}}{#1}}

\newif\if@cghi
\def\cite{\@cghitrue\@ifnextchar [{\@tempswatrue
	\@citex}{\@tempswafalse\@citex[]}}
\def\citelow{\@cghifalse\@ifnextchar [{\@tempswatrue
	\@citex}{\@tempswafalse\@citex[]}}
\def\@cite#1#2{{\if@cghi{$\null^{#1}$}\else#1\fi\if@tempswa\typeout
	{IJCGA warning: optional citation argument 
	ignored: `#2'} \fi}}

\def\pmb#1{\setbox0=\hbox{#1}
	\kern-.025em\copy0\kern-\wd0
	\kern.05em\copy0\kern-\wd0
	\kern-.025em\raise.0433em\box0}

\def\fnt#1#2{\footnotetext{\kern-.3em
	{$^{\mbox{\scriptsize #1}}$}{#2}}}

\def\fpage#1{\begingroup
\voffset=.3in
\thispagestyle{empty}\begin{table}[b]\centerline{\footnotesize #1}
	\end{table}\endgroup}

\def\runninghead#1#2{\pagestyle{myheadings}
\markboth{{\protect\footnotesize\it{\quad #1}}\hfill}
{\hfill{\protect\footnotesize\it{#2\quad}}}}
\font\tenrm=cmr10
\font\tenit=cmti10 
\font\tenbf=cmbx10
\font\bfit=cmbxti10 at 10pt
\font\ninerm=cmr9
\font\nineit=cmti9

\font\eightrm=cmr8

\def\Journal#1#2#3#4{{#1} {\bf #2}, #3 (#4)}

\def\PRL{{\bibit Phys. Rev. Lett.}}
\def\JHEP{{\bibit JHEP} }
\def\NP{{\bibit Nucl. Phys.}}
\def\PL{{\bibit Phys. Lett.}}
\def\PR{{\bibit Phys. Rev.}}

\def\JMP{{\bibit J. Math. Phys.}}
\def\ATMP{{\bibit Adv. Theor. Math. Phys.}}

\def\dalemb#1#2{{\vbox{\hrule height .#2pt
        \hbox{\vrule width.#2pt height#1pt \kern#1pt
                \vrule width.#2pt}
        \hrule height.#2pt}}}

  \let\q=\theta  
  \let\n=\nu

\let\C=\Chi      
\let\la=\label  
 \def\bd{\begin{document}} \def\ed{\end{document}}
\def\ds{\documentstyle} \let\fr=\frac \let\bl=\bigl \let\br=\bigr
\let\Br=\Bigr \let\Bl=\Bigl 
\let\bm=\bibitem
\let\na=\nabla
\let\pa=\partial \let\ov=\overline
\def\ie{{\it i.e.\ }} 
\newcommand{\be}{\begin{equation}} 
\newcommand{\ee}{\end{equation}} 
\def\ba{\begin{array}}
\def\ea{\end{array}}
\def\ft#1#2{{\textstyle{{\scriptstyle #1}\over {\scriptstyle #2}}}}
\def\fft#1#2{{#1 \over #2}}
\def\del{\partial}
\def\sst#1{{\scriptscriptstyle #1}}
\def\oneone{\rlap 1\mkern4mu{\rm l}}
\def\td{\tilde}
\def\wtd{\widetilde}
\def\im{{\rm i}}
\def\bog{Bogomol'nyi\ }
\def\q{{\tilde q}}
\def\hast{{\hat\ast}}
\def\0{{\sst{(0)}}}
\def\1{{\sst{(1)}}}
\def\2{{\sst{(2)}}}
\def\3{{\sst{(3)}}}
\def\4{{\sst{(4)}}}
\def\5{{\sst{(5)}}}
\def\6{{\sst{(6)}}}
\def\7{{\sst{(7)}}}
\def\8{{\sst{(8)}}}
\def\n{{\sst{(n)}}}
\newcommand{\w}[1]{\\[0.#1cm]}
\def\hA{\hat{\cal A}}
\def\ns{{\sst {\rm NS}}}
\def\rr{{\sst {\rm RR}}}
\def\tH{{\widetilde H}}
\def\tB{{\widetilde B}}
\def\cA{{\cal A}}
\def\cF{{\cal F}}
\def\tF{{\wtd F}}
\def\v{{{\cal V}}}
\def\Z{\rlap{\sf Z}\mkern3mu{\sf Z}}
\def\ep{{\epsilon}}
\def\IIA{{\rm IIA}}
\def\IIB{{\rm IIB}}
\def\ads{{\rm AdS}}
\def\R{\rlap{\rm I}\mkern3mu{\rm R}}
\def\ua{\underline{\alpha}}
\def\ub{\underline{\phantom{\alpha}}\!\!\!\beta}
\def\uc{\underline{\phantom{\alpha}}\!\!\!\gamma}
\def\um{\underline{\mu}}
\def\ud{\underline\delta}
\def\ue{\underline\epsilon}
\def\una{\underline a}
\def\unA{\underline A}
\def\unb{\underline b}
\def\unB{\underline B}
\def\unc{\underline c}
\def\unC{\underline C}
\def\und{\underline d}
\def\unD{\underline D}
\def\une{\underline e}
\def\unE{\underline E}
\def\unf{\underline{\phantom{e}}\!\!\!\! f}
\def\unF{\underline F}
\def\ung{\underline g}
\def\unm{\underline m}
\def\unM{\underline M}
\def\unn{\underline n}
\def\unN{\underline N}
\def\unp{\underline{\phantom{a}}\!\!\! p}
\def\unP{\underline P}
\def\unH{\underline{H}}
\def\unF{\underline{F}}
\def\unT{\underline{T}}
\def\ovA{\overline{A}}
\def\ovB{\overline{B}}
\def\uC{{\underline C}}
\def\ns{\normalsize}
\def\vs{\vspace{-0.25cm}}
\def\se{\;\;=\;\;}
\def\de{\;\;:=\;\;}
\def\cF{{\cal F}}
\def\cH{{\cal H}}
\def\cK{{\cal K}}

\def\cE{{\cal E}}
\def\tr{{\rm tr}}
\def\bC{{\bar \C}}

\newcommand{\bea}{\begin{eqnarray}} 
\newcommand{\eea}{\end{eqnarray}} 
\newcommand{\ra}{\rightarrow}
\newcommand{\Tr}{{\rm Tr} } 

\def\qed{\hbox{${\vcenter{\vbox{			
   \hrule height 0.4pt\hbox{\vrule width 0.4pt height 6pt
   \kern5pt\vrule width 0.4pt}\hrule height 0.4pt}}}$}}

\renewcommand{\thefootnote}{\fnsymbol{footnote}}	

\begin{document}
\begin{flushright}

\hfill\ \ \ {MCTP-00-01}\ \ \
 
{hep-th/0012249}\\
\end{flushright}

\runninghead{State of the Unification Address}
{State of the Unification Address}

\normalsize\textlineskip
\thispagestyle{empty}
\setcounter{page}{1}

\copyrightheading{}			

\vspace*{0.88truein}

\fpage{1}
\centerline{\bf STATE OF THE UNIFICATION ADDRESS\footnote{
Plenary talk delivered at The Division of Particles and Fields Meeting of 
The American Physical Society, August 9-12 2000, Ohio State University. 
Research supported in part by DOE Grant DE-FG02-95ER40899.}}
\vspace*{0.37truein}
\centerline{\footnotesize M. J. DUFF\footnote{mduff@umich.edu}}
\vspace*{0.015truein}
\centerline{\footnotesize\it Michigan Center for Theoretical Physics}
\baselineskip=10pt
\centerline{\footnotesize\it Randall Laboratory, Department of Physics,
University of Michigan}
\centerline{\footnotesize\it Ann Arbor, MI 48109--1120, USA}

\vspace*{0.21truein}
\abstracts{After reviewing how M-theory subsumes string theory,
I report on some new and interesting developments, focusing on the 
``brane-world'': circumventing no-go theorems for supersymmetric 
brane-worlds, complementarity of the Maldacena and Randall-Sundrum 
pictures; self-tuning of the cosmological constant.  I conclude with 
the top ten unsolved problems.}{}{}


\setcounter{footnote}{0}
\renewcommand{\thefootnote}{\alph{footnote}}

\vspace*{1pt}\textlineskip	
\section{Introduction}		
\vspace*{-0.5pt}
\noindent
Mr.~Chairman, members of DPF, honored guests, my fellow physicists: Today,
I have the honor of reporting on the State of the Unification. By which
I mean current developments in string theory and M-theory. Back in the
Fall of 99 when my colleagues and I at the University of Michigan began
planning for the organization of the Strings 2000
conference\footnote{http://feynman.physics.lsa.umich.edu/strings2000/}~ held
in July, we were concerned, as conference organizers often are, that perhaps
too little would have happened in the subject since Strings 99. In fact, we
need not have worried. Although there was nothing to rival the magnitude of
1984 String Revolution\cite{Schwarz1} or the 1995 M-theory
revolution,\cite{Duff0} the field is nevertheless very healthy with many
new and unexpected areas of progress. These include topics which go by
the names of: (1) non-commutative field theory, (2) open-membrane (OM) 
theory, (3) K-theory, (4) tachyon condensation, (5) strongly coupled gravity, 
(6) supersymmetric brane-worlds and how no-go theorems are circumvented,
(7) complementarity of the Maldacena and Randall-Sundrum pictures (8)
self-tuning of the cosmological constant.  My job is to summarize 
these new developments, but in view of the time constraints and in view of
the recent excellent reviews\cite{Schwarz2,Schwarz3} on topics (1)-(5),
I shall focus mainly on topics (6)-(8) which have the common theme of the
``brane-world.'' Bearing in mind the audience at this DPF meeting, these
are also the areas which seem closest to phenomenology.\footnote{``Not
close enough,'' I hear some of you say. I share your pain.}~ First,
however, it is necessary to recall where we are at the moment.

\newpage

\section{The story so far}
\subsection{M-theory and dualities}
\noindent
Not so long ago it was widely believed that there were five different 
superstring 
theories each competing for the title of ``Theory of everything,''
that all-embracing theory that describes all physical phenomena. See 
Table \ref{Strings}. 

Moreover, on the $(d,D)$ ``branescan'' of 
supersymmetric extended objects with $d$ worldvolume dimensions 
moving in a spacetime 
of $D$ dimensions, all these theories occupied the same 
$(d=2,D=10)$ slot. See table \ref{branescan}. The orthodox wisdom was 
that while $(d=2,D=10)$ 
was the Theory of Everything, the other branes on the scan were 
Theories of Nothing. 
All that has now changed. We now know that there are not five 
different theories at all but, together with $D=11$ supergravity, they 
form 
merely six different corners of a deeper, unique and more profound 
theory called ``M-theory'' where M stand for Magic, Mystery or 
Membrane. M-theory involves all of the other branes on the branescan, 
in particular 
the eleven-dimensional membrane $(d=3,D=11)$ and eleven-dimensional 
fivebrane $(d=6,D=11)$, thus 
resolving the mystery of why strings stop at ten dimensions while 
supersymmetry allows eleven.\cite{Duff0}

Although we can glimpse various corners of $M$-theory, the big 
picture 
still eludes us. Uncompactified $M$-theory has no dimensionless 
parameters, 
which is good from the uniqueness point of view but makes ordinary 
perturbation 
theory impossible since there are no small coupling constants to 
provide the expansion parameters.  A low 
energy, $E$, expansion is possible in powers of $E/M_{P}¥$, with 
$M_{P}¥$ the 
Planck mass, and leads to the familiar $D=11$ supergravity plus 
corrections of higher powers in the curvature. Figuring out what 
governs these corrections would go a long way in pinning down what 
$M$-theory really is. 

\begin{table}
\begin{center}
\vspace{.2in}
\begin{tabular}{l|c|c|c}
&\textbf{Gauge Group} & \textbf{Chiral?} & \textbf{Supersymmetry 
charges}\\ \hline 
\textbf{Type I}& $SO(32)$ & yes & 16\\ \hline 
\textbf{Type IIA} & $U(1)$ & no & 32\\ \hline
\textbf{Type IIB} & -- & yes & 32\\ \hline 
\textbf{Heterotic} & $E_8\times E_8$ & yes & 16\\ \hline 
\textbf{Heterotic} & $SO(32)$ & yes &16 \\
\hline
\end{tabular}
\end{center}
\caption{The Five Superstring Theories}
\la{Strings}
\end{table}

Why, therefore, do we place so much trust in a theory we cannot even 
define? First we know that its equations (though not in general its 
vacua) 
have the maximal number of 32 supersymmetry charges. This is a 
powerful 
constraint and provides many ``What else can it be?'' arguments in 
guessing what the theory looks like when compactified to $D<11$ 
dimensions. For example, when $M$-theory is compactified on a circle 
$S^{1}¥$ of radius $R_{11}¥$, it leads to the Type $IIA$ string, with 
string 
coupling constant $g_{s}$ given by
\be
g_{s}=R_{11}¥^{3/2}¥   
\ee
We recover the weak coupling regime only when $R_{11}¥\rightarrow 0$, 
which 
explains the earlier illusion that the theory is defined in 
$D=10$. Similarly, if we compactify on a line segment (known 
technically as $S^{1}¥/Z_{2}$) we recover the $E_{8}¥ \times E_{8}$ 
heterotic string.  Moreover, although the corners of M-theory we 
understand best 
correspond to the weakly coupled, perturbative, regimes where the 
theory can be approximated by a string theory, they are 
related to one another by a web of dualities, some of which are 
rigorously established and some of which are still conjectural but 
eminently plausible. For example, if we further compactify Type $IIA$ 
string on a 
circle of radius 
$R$, we can show rigorously that it is equivalent to the Type $IIB$ 
string 
compactified on a circle or radius $1/R$. If we do the same thing for 
the $E_{8}¥ \times E_{8}$ heterotic string we recover the $SO(32)$ 
heterotic string. These well-established relationships which remain 
within the 
weak coupling regimes are called {\it T-dualities}. The 
name {\it $S$-dualities} refers to the less well-established 
strong/weak coupling 
relationships. For example, the $SO(32)$ heterotic string is believed 
to be $S$-dual to the $SO(32)$ Type $I$ string, and the Type $IIB$ 
string to be self-$S$-dual. If we compactify more dimensions, other 
dualities can appear. For example, the heterotic string compactified on a 
six-dimensional torus $T^{6}¥$ is also believed to be self-$S$-dual. 
There is also the phenomenon of {\it duality of dualities} 
by which the $T$-duality of one theory is the $S$-duality of another. 
When M-theory is compactified on $T^{n}$, these $S$ and $T$ dualities 
are combined into what are termed $U$-dualities. 
All the consistency checks we have been able to 
think of (and after 5 years there dozens of them) have worked and 
convinced us that all these dualities are in fact valid. Of course we 
can compactify M-theory on more complicated manifolds such as the 
four-dimensional $K3$ or the six-dimensional Calabi-Yau spaces and 
these lead to a bewildering array of other dualities. For example: the 
heterotic string on $T^{4}$ is 
dual to the Type $II$ string on $K3$; the heterotic 
string on $T^{6}$ is dual to the    
the Type $II$ string on Calabi-Yau; the Type IIA string on a 
Calabi-Yau 
manifold is dual to the Type IIB string on the mirror Calabi-Yau 
manifold.
These more complicated compactifications lead to many more parameters 
in the theory, known to the mathematicians as {\it moduli}, but in 
physical uncompactified spacetime have the interpretation as 
expectation values of scalar fields. Within string perturbation 
theory, these scalar fields have flat potentials and their 
expectation 
values are arbitrary. So deciding which topology Nature 
actually chooses and the values of the moduli within that topology is 
known as the {\it vacuum degeneracy problem}.

\subsection{Branes}

\begin{table}
$
\begin{array}{ccccccccccccccc}
~&D\uparrow&&&&&&&&&&&~\\
~&11&.&~&&S&&&T&&&&&~\\
~&10&.&V&S/V&V&V&V&S/V&V&V&V&V&~\\
~&9&.&S&&&&S&&&&&&~\\
~&8&.&~&&&S&&&&&&&~\\
~&7&.&~&&S&&&T&&&&&~\\
~&6&.&V&S/V&V&S/V&V&V&&&&&~\\
~&5&.&S&&S&&&&&&&&~\\
~&4&.&V&S/V&S/V&V&&&&&&&~\\
~&3&.&S/V&S/V&V&&&&&&&&~\\
~&2&.&S&&&&&&&&&&~\\
~&1&.&~&~&~&~&~&~&~&~&~&~&~\\
~&0&.&.&.&.&.&.&.&.&.&.&.&.~\\
~&~&0&1&2&3&4&5&6&7&8&9&10&11&d\rightarrow
\end{array}
$
\caption{The branescan, where $S$, $V$ and $T$ denote scalar, 
vector and
antisymmetric tensor multiplets.}
\la{branescan}
\end{table}

In the previous section we outlined how M-theory makes contact with 
and relates the previously known superstring theories, but as its 
name suggest, $M$-theory also relies heavily on membranes or more 
generally 
$p$-branes, extended objects with $p=d-1$ spatial dimensions (so a 
particle is 
a 0-brane, a string is a 1-brane, a membrane is a 2-brane and so on). 
In $D=4$, a charged 0-brane couples naturally to an Maxwell vector 
potential 
$A_{\mu}$, with field strength $F_{\mu\nu}¥$ and carries an electric 
charge
\be
Q \sim \int_{S^{2}¥}¥*F_{2}¥ 
\ee
and magnetic charge
\be
P \sim \int_{S^{2}¥}¥F_{2}¥ 
\ee
where $F_{2}¥$ is the Maxwell 2-form field strength, $*F_{2}¥$ is its 
2-form dual and $S^{2}¥$ is a 2-sphere surrounding the charge. This 
idea may be generalized 
to $p$-branes in $D$ dimensions.
A $p$-brane couples to $(p+1)$-form potential 
$A_{\mu_{1}¥\mu_{2}¥\ldots\mu_{p+1}¥}¥$ 
with $(p+2)$-form field strength 
$F_{\mu_{1}¥\mu_{2}¥\ldots\mu_{p+2}¥}¥$ 
and carries an ``electric'' charge per unit p-volume
\be
Q \sim \int_{S^{D-p-2}¥}¥*F_{D-p-2}¥ 
\ee
and ``magnetic'' charge per unit p-volume
\be
P \sim \int_{S^{p+2}¥}¥F_{p+2}¥ 
\ee
where $F_{p+2}¥$ is the $(p+2)$-form field strength, $*F_{D-p-2}¥$ 
its $(D-p-2)$-form dual and $S^{n}$ is an n-sphere surrounding the 
brane. A special role is played in $M$-theory by the BPS 
(Bogomolnyi-Prasad-Sommerfield) branes whose mass per unit 
$p$-volume, 
or tension $T$, is equal to the charge per unit $p$-volume
\be
T\sim Q
\ee
This formula may be generalized to the cases where the branes carry 
several electric and magnetic charges. The supersymmetric branes shown 
on the branescan are always BPS, although the converse is not true. 
M-theory also makes use of non-BPS and non-supersymmetric branes not 
shown on the branescan, but the supersymmetric ones do play a special role 
because they are guaranteed to be stable.

The letters S, V, and T on the branescan refer to scalar, vector and 
antisymmetric tensor supermultiplets of fields that propagate on the 
worldvolume of the brane. Historically, these points on the branescan 
were discovered in three different ways. The S branes were classified 
by writing down spacetime supersymmetric worldvolume actions that 
generalize the Green-Schwarz actions on the superstring
worldsheet.\cite{Achucarro} By contrast, the V and T branes were
shown to arise as 
soliton solutions\footnote{The 3-brane soliton of Type IIB 
supergravity was an early 
candidate for a `brane-world', firstly because of its 
dimensionality\cite{Horowitz,Dufflu} and secondly because gauge fields propagate 
on its worldvolume\cite{Dufflu}.  See section 3.2.} of the underlying supergravity theories\cite{Duffkhuri}.
However, the solitonic V branes found this way were bound 
by $p \leq 7$. The 8-brane and 9-brane slots were included on the scan only 
because they were allowed by supersymmetry.\cite{Duffkhuri} Subsequently, all 
the V-branes 
were given a new interpretation as Dirichlet p-branes, called D-branes, 
surfaces of 
dimension p on which open strings can end and which carry R-R 
(Ramond-Ramond) charge.\cite{Polchinski} The IIA theory has D-branes with $p=0,2,4,6,8$ 
and the IIB theory has D-branes with $p=1,3,5,7,9$. They are related to one another by 
T-duality. In terms of how their tensions depend on the string 
coupling $g_{s}¥$, the D-branes are 
mid-way between the fundamental (F) strings and the solitonic (S) 
fivebranes: 
\be
T_{F1}¥\sim m_{s}¥^{2}¥,~~~~T_{Dp}¥\sim \frac{m_{s}¥^{p+1}¥}{g_{s}¥}, 
~~~~T_{S5}¥\sim \frac{m_{s}¥^{6}¥}{g_{s}¥^{2}¥}
\ee
Since they are BPS, there is a no-force condition between the branes that allows 
us to have many branes of the same charge parallel to one another. The gauge group 
on a single D-brane is an abelian $U(1)$.  If we stack $N$ such branes on top of one 
another, the gauge group is the non-abelian $U(N)$. As we separate them this decomposes 
into its subgroups, so in fact there is a Higgs mechanism at work whereby the 
vacuum expectation values of the Higgs fields are related to the separation of the 
branes. For example the theory that lives on a stack of $N$ Type IIB $D3$ 
branes is a four-dimensional $U(N)$ $n=4$ super Yang-Mills theory. In 
the limit of large $N$ the geometry of this configuration tends to the 
product of five dimensional anti-de Sitter space and a five dimensional 
sphere, $AdS_{5}¥\times S^{5}¥$.   

In $D=11$, M-theory has two BPS branes, an electric 2-brane and its 
magnetic dual which is a 5-brane. Their tensions are related to each 
other and the Planck mass by
\be
T_{2}^{3} \sim T_{5} \sim M_{P}^{6}
\ee
if we stack $N$ such branes on top of one another, the M2-brane geometry tends 
in the large $N$ limit to $AdS_{4} \times 
S^{7}$ and the M5-brane geometry to $AdS_{7}¥ \times S^{4}¥$.
In addition there are two other objects in $D=11$, the plane wave and the 
Kaluza-Klein monopole, which though not branes are still BPS. When 
spacetime is 
compactified a $p$-brane may remain a $p$-brane or else 
become a $(p-k)$-brane if it wraps around $k$ of the compactified 
directions.
For example, the Type $IIA$ fundamental string emerges by wrapping 
the M2-brane around $S^{1}$ and shrinking its radius to zero, and the 
Type $IIA$ 4-brane emerges in a similar way from the $M5$-brane. 

\subsection{Spin-offs of M-theory} 
\noindent What do we now know with 
M-theory that we did not know with old-fashioned string theory?  Here 
are a few examples, references to which may be found in 
Ref.~\citelow{Duff0}.

1) Electric-magnetic (strong/weak coupling) duality in $D=4$ is a consequence of 
string/string duality in $D=6$ which in turn is a consequence of 
membrane/fivebrane duality in $D=11$. 

2) {\it Exact} electric-magnetic duality, first proposed for the maximally 
supersymmetric conformally invariant $n=4$ super Yang-Mills theory, has been 
extended to {\it effective} duality by Seiberg and Witten to 
non-conformal $n=2$ theories: the so-called Seiberg-Witten theory. 
This has been very successful in providing the first proofs of quark 
confinement (albeit in the as-yet-unphysical super QCD) and in 
generating new pure mathematics on the topology of four-manifolds. 
Seiberg-Witten theory and other $n=1$ dualities of Seiberg may, 
in their turn, be derived from M-theory.

3) Indeed, it seems likely that all supersymmetric quantum field 
theories with any gauge group, and their spontaneous symmetry 
breaking, admit a geometrical 
interpretation within M-theory as the worldvolume fields that propagate on 
the common intersection of stacks of p-branes wrapped around various 
cycles of the compactified dimensions, with the Higgs expectation 
values given by the brane separations.

4) In perturbative string theory, the vacuum degeneracy problems arises 
because there are billions of Calabi-Yau vacua which are distinct according 
to classical topology. Like higher-dimensional Swiss cheeses, each can have 
different number of $p$-dimensional holes. This results in many 
different kinds of four-dimensional gauge theories with different gauge 
groups, numbers of families and different choices of quark and lepton 
representations. Moreover, M-theory introduces new non-perturbative 
effects which allow many more possibilities, making the degeneracy 
problem apparently even worse. However, most (if not all) of these manifolds 
are in fact smoothly connected in M-theory by shrinking the $p$-branes that 
can wrap around the $p$-dimensional holes in the manifold and which 
appear as black holes in spacetime. As the wrapped-brane volume shrinks to zero, 
the black holes become massless and effect a smooth transition from one 
Calabi-Yau manifold to another. Although this does not yet cure the 
vacuum degeneracy problem, it puts it in a different light. The 
question is no longer why we live in one topology rather than another but 
why we live in one particular corner of the unique topology. This may 
well have a dynamical explanation.

5) Ever since the 1970's, when Hawking used macroscopic arguments to predict 
that black holes have an entropy equal to one quarter the area of their  
event horizon, a microscopic explanation has been lacking. But treating black 
holes as wrapped $p$-branes, together with the realization that Type II branes 
have a dual interpretation as Dirichlet branes, allows the first microscopic 
prediction in complete agreement with Hawking. The fact that M-theory 
is clearing up some long standing problems in quantum gravity gives us 
confidence that we are on the right track. 

6) It is known that the strengths of the four forces change with energy. In 
supersymmetric extensions of 
the standard model, one finds that the fine structure constants
$\alpha_3,\alpha_2,\alpha_1$ associated with the $SU(3) \times SU(2)
\times U(1)$  all meet at about $10^{16}$ GeV, entirely consistent with
the idea of grand unification.  The strength of the dimensionless number
$\alpha_G=GE^2$, where $G$ is Newton's constant and $E$ is the energy, also
almost meets the other three, but not quite. This near miss has been a
source of great interest, but also frustration.  However, in a universe of
the kind envisioned by Witten, spacetime is
approximately a narrow five dimensional layer bounded by four-dimensional
walls. The particles of the standard model live on the walls but gravity
lives in the five-dimensional bulk. As a result, it is possible to  choose
the size of this fifth dimension so that all four forces meet at this
common scale.  Note that this is much less than the Planck scale of
$10^{19}$ GeV, so gravitational effects may be much closer in energy than
we previously thought; a result that would have all kinds of cosmological
consequences.

So what is $M$-theory?

There is still no definitive answer to this question, although 
several different proposals have been made. By far the most popular 
is M(atrix) theory\cite{Banks}. The matrix models of $M$-theory are 
$U(N)$ supersymmetric gauge quantum mechanical models with $16$ 
supersymmetries.  Such models are also interpretable as the effective 
action of $N$ coincident Dirichlet $0$-branes. 

The theory begins by compactifying the eleventh dimension on a circle of 
radius $R$, so that the longitudinal momentum is quantized in units 
of $1/R$ with total $P_{L}$ $N/R$ with $N \rightarrow \infty$. The 
theory is {\it holographic} in that it contains only 
degrees of freedom which carry the smallest unit of longitudinal 
momentum, other states being composites of these fundamental states.  
This is, of course entirely consistent with their identification with 
the Kaluza-Klein modes. It is convenient to describe these $N$ degrees of 
freedom as $N\times 
N$ matrices.  When these matrices commute, their simultaneous eigenvalues are 
the positions of the $0$-branes in the conventional sense. That they will 
in general be non-commuting, however, suggests that to properly 
understand $M$-theory, we must entertain the idea of a fuzzy spacetime 
in which spacetime coordinates are described by non-commuting matrices. 
In any event, this matrix approach has had success in reproducing many 
of the expected properties of $M$-theory such as $D=11$ Lorentz 
covariance, $D=11$ supergravity as the low-energy limit, and the 
existence of membranes and fivebranes. 

It was further proposed that when 
compactified on $T^{d-1}$, the quantum mechanical model should be replaced 
by an $d$-dimensional $U(N)$ Yang-Mills field theory defined on 
the dual torus ${\tilde T}^{d-1}$.  Another test of this M(atrix) approach, then, 
is that it should explain the $U$-dualities. For 
$d=4$, for example, this group is $SL(3,Z) \times SL(2,Z)$. The 
$SL(3,Z)$ just comes from the modular group of $T^{3}$ whereas the 
$SL(2,Z)$ is the electric/magnetic duality group of four-dimensional 
$n=4$ Yang-Mills. For $d>4$, however, this picture 
looks suspicious because the corresponding gauge theory becomes 
non-renormalizable and the full $U$-duality group has still escaped    
explanation. There have been speculations on what 
compactified $M$-theory might be, including a revival of the old proposal 
that it is really M(embrane)theory. In other words, perhaps 
$D=11$ supergravity together with its BPS configurations: plane wave, 
membrane, fivebrane, KK monopole and the $D=11$ embedding of the 
Type $IIA$ eightbrane, are all there is to $M$-theory and that we 
need look no further for new degrees of freedom, but only for a new 
non-perturbative quantization scheme. At the time of 
writing this is still being hotly debated.

What seems certain, however, is that M-theory is not a string theory. 
It can be approximated by a string theory only in certain peculiar 
corners of parameter space. So ``string phenomenology'' will become an 
oxymoron unless, for some as yet unknown reason, our universe happens 
to occupy one of these corners.

\subsection{AdS/CFT and the brane-world}
\noindent
The year 1998 marked a renaissance in anti de-Sitter space (AdS) 
brought about by Maldacena's conjectured duality between physics in the 
bulk of $AdS$ and a conformal field theory on its boundary.\cite{Maldacena}
For example, $M$-theory on $AdS_{4}¥\times S^{7}$ is dual to a 
non-abelian $(n=8,d=3)$ superconformal theory, Type $IIB$ string theory on 
$AdS_{5}¥\times S^{5}$ is dual to a $(n=4,d=4)$ $U(N)$ super Yang-Mills  
theory and $M$-theory on $AdS_{7}¥\times S^{4}$ is dual to a non-abelian 
$((n_+,n_-)=(2,0),d=6)$ conformal theory. In particular, as has been spelled 
out most clearly in the $d=4$ $U(N)$ Yang-Mills case, there is seen to be a 
correspondence between the Kaluza-Klein mass spectrum in the bulk and the 
conformal dimension of operators on the boundary.\cite{Gubser,Wittenads}
We note that, by choosing Poincar\'e coordinates on AdS$_5$, the metric may be 
written as
\begin{equation}
ds^2=e^{-2y/L}(dx^\mu)^2+dy^2,
\end{equation}
where $x^{\mu}$, $(\mu=0,1,2,3)$, are the four-dimensional brane coordinates. 
In this case the superconformal Yang-Mills theory is taken to reside at the
boundary $y\to-\infty$. The AdS length scale $L$ is given by
\be
L^{4} =4\pi\alpha'^{2}¥(g_{YM}^{2} N) 
\ee
The string coupling $g_{s}$ and the Yang-Mills coupling $g_{YM}$ are related by  
\be
g_{s}¥=g_{YM}¥^{2}
\ee
The full quantum string theory on this spacetime is difficult to deal 
with, but we can approximate it by classical Type IIB supergravity provided 
\be
 L^{2}¥>>{\alpha}' 
\ee
so that stringy correction to supergravity are small, and that 
$g_{s}<<1$ or
\be
N\rightarrow \infty
\ee
so that loop corrections can be neglected. There is now overwhelming 
evidence in favor of this correspondence and it allows us to calculate 
previously uncalculable strong coupling effects in the gauge theory 
starting from classical supergravity.  Models of this kind, where a bulk 
theory with gravity is equivalent to a boundary theory without 
gravity, have also been advocated by `t Hooft\cite{thooft} and by 
Susskind\cite{Susskind1} who call them {\it holographic} theories. 
Many theorists are understandably excited about the AdS/CFT 
correspondence because of what it can teach us about non-perturbative 
QCD. In my opinion, however, this is, in a sense, a 
diversion from the really fundamental question: What is $M$-theory?
So my hope is that this will be a two-way process and that 
superconformal field theories will also teach us more about $M$-theory.

The Randall-Sundrum mechanism\cite{Randall} also involves AdS but was originally
motivated, not via the decoupling of gravity from D3-branes,
but rather as a possible mechanism for evading Kaluza-Klein compactification
by localizing gravity in the presence of an uncompactified extra dimension.
This was accomplished by inserting a positive tension 3-brane (representing
our spacetime) into AdS$_5$.  In terms of the Poincar\'e patch of AdS$_5$
given above, this corresponds to removing the region $y<0$, and either
joining on a second partial copy of AdS$_5$, or leaving the brane at the end
of a single patch of AdS$_5$.  In either case
the resulting Randall-Sundrum metric is given by
\begin{equation}
ds^2=e^{-2|y|/L}(dx^\mu)^2+dy^2,
\end{equation}
where $y\in(-\infty,\infty)$ or $y\in[0,\infty)$ for a `two-sided' or
`one-sided' Randall-Sundrum brane respectively.

The similarity of these two scenarios led to the notion that they
are in fact closely tied together.  To make this connection clear,
consider the one-sided Randall-Sundrum brane.  By introducing a boundary
in AdS$_5$ at $y=0$, this model is conjectured to be dual to a cutoff CFT
coupled to gravity, with $y=0$, the location of the Randall-Sundrum
brane, providing the UV cutoff.  This extended version of the Maldacena
conjecture\cite{wittensusskind}
then reduces to the standard AdS/CFT duality as the boundary
is pushed off to $y\to-\infty$, whereupon the cutoff is removed and gravity
becomes completely decoupled.  Note in particular that this connection
involves a single CFT at the boundary of a single patch of AdS$_5$.  For the
case of a brane sitting between two patches of AdS$_5$, one would instead
require two copies of the CFT, one for each of the patches.  A crucial 
test of this assumed complementarity of the Maldacena and 
Randall-Sundrum pictures is that both should yield the same 
corrections to Newton's law.  See section 3.3.

A third development in the brane-world has been the idea that the 
extra dimensions are compact but much larger than the conventional 
Planck sized dimensions in traditional Kaluza-Klein theories\cite{Dimopoulos}.
This is possible if the standard model fields are confined to the $d=4$ brane 
with only gravity propagating in the $d>4$ bulk. We shall return to 
this possibility in section 3.4. 

\newpage

\section{Developments on the Brane-World}
\la{brane}
\subsection{No-go theorems for supersymmetry}
\la{nogo}
\noindent
If we are to give a ``top-down'' justification of the Randall-Sundrum brane-world by 
embedding it in string theory or M-theory, it is desirable that the R-S picture 
be consistent with supersymmetry. Indeed, such a supersymmetric 
brane-world is necessary if the Maldacena and Randall-Sundrum (R-S) pictures are 
to stand any chance of being complementary. At first, however, this seemed to be 
problematical and several papers appeared in the literature suggesting 
that R-S could not be supersymmetric. Some of these no-go theorems listed below 
are exactly as they appeared; with others I have taken the liberty of setting up 
the straw man so as more effectively to knock him down.

1) {\bf R-S branes cannot be SUSY because massless supergravity scalars give 
kink-up and not kink-down potentials which do not bind gravity to the 
brane.}

2) {\bf R-S branes cannot be SUSY because their tension is not that of a BPS
brane.}

3) {\bf R-S branes cannot be SUSY because $\delta$-functions are 
incompatible with susy transformation rules.}

4) {\bf R-S branes cannot be SUSY because the photon superpartners of the 
graviton cannot be bound to the brane.}

\subsection{Yes-go theorems for Supersymmetry of the brane-world}
\la{yesgo}
\noindent
In fact, the domain-wall solution of Bremer et al\cite{Bremer} 
provides a supersymmetric Type IIB Randall-Sundrum 
realization\cite{DLS}. See also Refs.~\citelow{clp} and 
\citelow{deAlwis}.
So it is instructive to see how the no-go theorems are circumvented:

1) {\bf The required supergravity scalar is massive, being the breathing 
mode of the $S^{5}$ compactification\cite{DLS}.} So the negative 
conclusions about massless scalars in Refs.~\citelow{Kalloshlinde,Behrndt},
while correct, are not relevant.
 
2) {\bf The tension comes from two sources: the BPS D3-branes and the
kink\cite{Cvetic3}.} So the observation of Ref.~\citelow{Krauss} that
the D3 brane tension is only 2/3 of the R-S tension, while correct,
is not relevant.

3) {\bf The sign flip of the coupling constant across the brane removes 
the $\delta$-functions in the supersymmetry transformation
rules\cite{Ghergetta,Ortin,Kallosh,DLS,Sati}.} So the problems raised by 
Ref.~\citelow{Bagger}, while correct, are not relevant.

4) {\bf Photons can be bound to the brane but their bulk origin is not 
Maxwell's equations but rather odd-dimensional self-duality equations \cite{Sabra,lupope}.} 
So the result of Ref.~\citelow{Hodge}, that photons obeying Maxwell's 
equations in the bulk cannot be bound to the brane, while correct, is 
not relevant\footnote{The authors of 
Refs.~\citelow{Bajc,Pomarol,Kaloper} showed that, treated as test particles, 
Maxwell photons could be bound to the brane but their charge would be 
screened.  However, the combined bulk Einstein-Maxwell equations rule 
out photons on the brane altogether \cite{Hodge}.}

An entirely different question is whether a {\it smooth} domain wall 
can provide a supersymmetric Randall-Sundrum realization, and here 
ordinary supergravity seems to fail requiring some kind of higher 
derivative and presumably stringy corrections.\cite{Kalloshlinde,Maldacena1}
 
\subsection{Complementarity of the Maldacena and Randall-Sundrum pictures}
\la{complement}
\noindent
In his 1972 PhD thesis under Abdus Salam, the author showed that, 
when one-loop quantum corrections to the graviton propagator are taken into account, 
the inverse square 
$r^{-2}$ behavior of Newton's gravity force law receives an $r^{-4}$ 
correction 
whose coefficient depends on the number and spins of the elementary 
particles.\cite{Duff1,Duff2} Specifically, the potential looks like 
\be
V(r)= \frac{GM}{r}\biggl
(1+\frac{\alpha G}{r^{2}} \biggr),
\la{Newton}
\ee
where $G$ is the four-dimensional Newton's constant, ${\hbar}=c=1$ and
$\alpha$ is a purely numerical coefficient given, in the case of spins
$s\leq1$, by $45 \pi \alpha=12N_{1}+3N_{1/2}+N_{0}$, where $N_{s}$ are
the numbers of particle species of spin $s$ going around the loop.

Now fast-forward to 1999 when Randall and Sundrum proposed that 
our four-dimensional world is a 3-brane embedded in an infinite 
five-dimensional universe. Gravity reaches out into the 
five-dimensional bulk but the other forces are confined to the four-dimensional 
brane. Contrary to expectation, they showed that an inverse square $r^{-2}$ 
law for gravity is still possible but 
with an $r^{-4}$ correction coming from the massive Kaluza-Klein 
modes whose coefficient depends on the bulk 
cosmological constant. Their potential looks like
\be
V(r)= \frac{GM}{r}\biggl(1+\frac{2L^{2}}{3 r^{2}} \biggr).
\la{RS}
\ee
where $L$ is the radius of AdS$_{5}$.  Since (\ref{Newton}) was the result of a 
four-dimensional quantum 
calculation and (\ref{RS}) the result of a five-dimensional 
classical calculation, they seem superficially completely unrelated.
However, Ref.~\citelow{Duffliu} invokes the  
AdS/CFT correspondence 
of Maldacena to 
demonstrate that the two are in fact completely 
equivalent ways of describing the same physics.  From (\ref{Newton}),
we see that the contribution of a single $n=4$ $U(N)$ Yang-Mills CFT, with
$(N_{1},N_{1/2},N_{0})=(N^{2},4N^{2},6N^{2})$, is
\be
V(r)= \frac{GM}{r}\biggl
(1+\frac{2N^{2}G}{3\pi r^{2}} \biggr).
\la{AdS}
\ee
Using the AdS/CFT relation $N^{2}=\pi L^{3}/2G_{5}$  and the
one-sided brane-world relation $G=2G_{5}/L$,
where $G_{5}$ is the five-dimensional
Newton's constant, we obtain 
exactly (\ref{RS}).

As discussed in the August 2000 edition of Scientific
American\cite{Dimopoulos}, experimental 
tests of deviations from Newton's inverse square law are currently under way.

\newpage

\subsection{Self-tuning of the cosmological constant}
\label{cosmo}
\noindent
Imagining that our universe is a brane in a higher dimensional spacetime 
changes the nature of the cosmological constant problem. We must now 
explain why the four-dimensional cosmological constant $\Lambda_{4}¥$ 
is small, without unacceptable fine-tuning, as opposed to the bulk 
cosmological constant $\Lambda_{5}¥$. An interesting recent development is 
the following ``self-tuning'' idea.

Once again, we start with a five-dimensional bulk theory with a scalar 
field $\phi$ coupled to a four-dimensional brane source via an action:
\begin{equation}
\int d^{4}¥x \sqrt{-g}Ve^{b\phi}¥
\end{equation}
where $V$ and $b$ are constants. Let us allow Poincare supersymmetry 
in the bulk while the theory on the brane breaks supersymmetry.  An 
example of a mechanism for supersymmetry breaking on the brane may be 
found in Ref.~\citelow{Antoniadis}.  Hence, to first approximation, 
the bulk has vanishing cosmological constant.  This has the important 
consequence that the bulk interactions are invariant under a shift in 
$\phi$:
\begin{equation}
\phi \rightarrow \phi + constant
\la{shift}
\end{equation}
The authors of Refs.~\citelow{Arkani} and \citelow{Kachru} were able to find 
3-brane solutions of the combined bulk-brane equations for which the 
four-dimensional cosmological constant vanishes. As usual, the question 
now is why it should remain zero after supersymmetry breaking and radiative 
corrections of the ``standard model'' theory on the brane. Normally we 
would expect huge corrections to $\Lambda_{4}$ of order $M_{S}¥^{4}¥$ where 
$M_{S}¥$ is the supersymmetry-breaking scale. 

The crucial observation is that, as a result of (\ref{shift}), given a 
brane solution with one value of $V$, there exist another for any 
value, since shifts in $V$ can be absorbed by shifts in $\phi$, owing 
to the combination $Ve^{b\phi}¥$.  Hence one can effectively absorb any 
cosmological constant generated by standard model physics, and keep 
$\Lambda_{4}=0$ to lowest order.  (The absence of a bulk cosmologcical 
constant makes this scenario different from that of sections 3.1, 3.2 
and 3.3.  However, there also exist more generalized models which 
permit a self-tuning with a non-zero bulk potential\cite{Csaki}.)

Of course, the interactions between bulk and brane fields will result 
in supersymmetry breaking on the brane eventually feeding back to the 
bulk and spoiling the $\Lambda_{5}=0¥$ starting point. However, the 
resulting corrections to $\Lambda_{4}¥$ 
appear as a power series in $M_{S}¥/M_{5}¥$, which will be about 
$10^{-5}$, if we take $M_{S} \sim TeV$ and $M_{4} \sim 10^{19}¥ 
GeV$. This is still much bigger than the experimental bound and so 
this is by no means the final answer to the cosmological constant 
problem. There are other possible objections to these models that 
might be raised. However, the main culprit  $M_{S}¥^{4}¥$ has at least 
been eliminated, and this may point the way to a new solution.

\newpage
\subsection{Five versus eleven}
\noindent
As we have seen M-theory requires eleven dimensions, whereas if the 
brane-world picture is correct, we really need only five with the other six going 
along for the ride. Why should Nature behave like this? The only  good 
answer to this question I could find is in Mother Goose's  Nursery 
Rhymes:

{\it Nature requires five,

Custom allows seven,

Idleness takes Nine,

And Wickedness Eleven.}

\section{Top Ten Unsolved Problems}
\noindent
In 1900 the world-renowned mathematician David Hilbert presented 
twenty-three problems at the International Congress of Mathematicians 
in Paris. 
These problems have inspired mathematicians throughout the last 
century. 
As a piece of millennial madness, all participants of the Strings 
2000 Conference 
were invited to help formulate the ten most important unsolved 
problems in fundamental
physics. Each participant was allowed to submit one candidate problem 
for consideration. 
To qualify, the problem must not only have been important but also 
well-defined and stated in a clear way. 
The best 10 problems were selected at the end of the conference by a 
selection panel consisting of 
David Gross, Edward Witten and myself. The results in no particular 
order are:

\begin{enumerate}

\item[(1)]{\it Are all the (measurable) dimensionless parameters that 
characterize the physical universe calculable in principle or are 
some merely determined by historical or quantum mechanical
              accident and uncalculable?} 
              
              David Gross, Institute for Theoretical Physics, 
University of California, Santa Barbara.

\item[(2)]{\it How can quantum gravity help explain the origin of the 
universe?} 
              
              Edward Witten, California Institute of Technology and 
Institute for Advanced Study, Princeton. 

\item[(3)]{\it What is the lifetime of the proton and how do we 
understand it?} 
              
              Steve Gubser, Princeton University and California 
Institute of Technology. 

\item[(4)]{\it Is Nature supersymmetric, and if so, how is 
supersymmetry broken?} 
              
              Sergio Ferrara, CERN European Laboratory of Particle 
Physics; Gordon Kane, University of Michigan.

\item[(5)]{\it Why does the universe appear to have one time and 
three space dimensions?} 
              
              Shamit Kachru, University of California, Berkeley; 
Sunil 
              Mukhi, Tata Institute of Fundamental Research; Hiroshi 
Ooguri, California Institute of Technology. 

\item[(6)]{\it Why does the cosmological constant have the value that 
it has, is it zero and is it really constant?} 
              
              Andrew Chamblin, Massachusetts Institute of 
Technology; Renata Kallosh, Stanford University. 

\item[(7)]{\it What are the fundamental degrees of freedom of 
M-theory (the theory whose low-energy limit is eleven-dimensional 
supergravity and which subsumes the five consistent
              superstring theories) and does the theory describe 
Nature? }
              
              Louise Dolan, University of North Carolina, Chapel 
              Hill; Annamaria Sinkovics, Spinoza Institute; Billy \& 
Linda 
              Rose, San Antonio College. 

\item[(8)]{\it What is the resolution of the black hole information 
paradox?} 
              
              Tibra Ali, Department of Applied Mathematics and 
Theoretical hPhysics, Cambridge; Samir Mathur, Ohio State 
University.

\item[(9)]{\it What physics explains the enormous disparity between 
the gravitational scale and the typical mass scale of the elementary 
particles? }
              
              Matt Strassler, Institute for Advanced Study, 
              Princeton. 

\item[(10)]{\it Can we quantitatively understand quark and gluon 
confinement in Quantum Chromodynamics and the existence of a mass 
gap? }
              
              Igor Klebanov, Princeton University; Oyvind Tafjord, 
McGill 
              University.
\end{enumerate}

\nonumsection{Acknowledgments}
\noindent
I am grateful to my collaborators Mirjam Cvetic, James T. Liu, Hong Lu, 
Chris Pope, Wafic Sabra and Kelly Stelle.

\nonumsection{References}

\end{document}